\documentclass[aps,pra,reprint,amsmath,amssymb,bibnotes,twocolumn]{revtex4-2}
\usepackage{xcolor,color,graphicx,float,hyperref,amsmath}

\begin{document}
\draft
\def\ds{\displaystyle}

\title{Nonlinear skin modes and fixed points}
\author{C. Yuce}
\affiliation{Department of Physics, Eskişehir Technical University, Eskişehir 26555, Türkiye}
\email{cyuce@eskisehir.edu.tr}

\date{\today}
\begin{abstract} We investigate a one-dimensional tight-binding lattice with asymmetrical couplings and various type of nonlinearities to study nonlinear non-Hermitian skin effect. Our focus is on the exploration of nonlinear skin modes through a fixed-point perspective. The nonlinear interactions are shown to have no impact on the spectral region in the semi-infinite system; however, they induce considerable changes when boundaries are present. The spectrum under open boundary conditions is found not to be a subset of the corresponding spectrum under the semi-infinite boundary conditions. We identify distinctive features of nonlinear skin modes, such as degeneracy, and power-energy discontinuity. Furthermore, we demonstrate that a family of localized modes that are neither skin nor scale-free localized modes is formed with the introduction of a coupling impurity. Additionally, we show that an impurity can induce discrete dark and anti-dark solitons.
\end{abstract}

\maketitle

\section{Introduction}
The non-Hermitian skin effect leads to a large number of localized modes at the boundaries of the lattice, which is not possible in its Hermitian counterpart \cite{wangin,sw1,slager,sw2,sw3}. This effect can change both the spectral and dynamical properties of a system, giving rise to intriguing dynamical effects such as topological funneling and edge bursts \cite{scifun,scifun2,in3}. Scale-free localization is another unique localization phenomena in non-Hermitian systems \cite{sfl3,sflee}. This type of localization is different from both skin and  Anderson localization as it implies size-dependent localization lengths \cite{sfl1,sfl2}. Scale-free localization challenges conventional understanding of localization concepts \cite{sflberg,sflmuro,sfl10,sfl9,sfl6,sflyeni,sfl8,sfl5,sflcem}. We note that even a single non-Hermitian defect in an otherwise Hermitian lattice can induce many localized modes whose localization lengths scale with system size \cite{wangin3}. \\
The non-Hermitian skin effect has been extensively studied, but only few papers extending this effect into the nonlinear regime have appeared in the literature 
\cite{nonlin1,gsdye,nonlin2,nonlin3,nonlin4,nonlin5,nonlin6,nonlin7,nlekle0,nlekle1,nlekle2,nonlin9}. For instance, we explored the nonlinear skin effect and constructed nonlinear skin modes, and discussed fractal structure \cite{nonlin1}. M. Ezawa further extended the understanding of this effect by investigating dynamical nonlinear higher-order skin effect and identified a novel topological phase \cite{nonlin2}. Experimental evidence of anomalous single-mode lasing, driven by the combined effects of nonlinearity and non-Hermiticity, was provided in \cite{nonlin3}. Recent studies have analyzed the nonlinear perturbations of systems with high-order exceptional points, uncovering skin discrete breathers and hierarchical power-law scaling behaviors \cite{nonlin4,nonlin5}. In addition to these developments, ref. \cite{nonlin6} have explored the emergence of insensitive edge solitons in non-Hermitian topological lattices, which exhibit robustness against perturbations. Reference \cite{nonlin7} introduced the concept of solitons with self-induced topological nonreciprocity. An experimental demonstration of the nonlinear non-Hermitian skin effect and skin solitons in an optical system with Kerr nonlinearity has recently been realized \cite{nlekle0}.\\
The paper investigates the unique properties and behaviors of nonlinear skin modes in a generalized nonlinear system by using the numerical shooting method. We show that fixed points provides us construct nonlinear skin modes and its spectral regions for a finite size lattice and semi-infinite lattice. We then outline the key similarities and differences between linear and nonlinear skin modes.  Furthermore, we show that a coupling impurity induces a family of localized modes that are neither skin nor scale free localized modes, and also a discrete dark and anti-dark solitons. 
\section{ Model}
We start with the equation for the complex field amplitude $\ds{\psi_n}$ with $n=1,2,...,N$, describing one-dimensional tight-binding nonlinear lattices with asymmetric couplings $J_{L}$ and $J_{R}$ 
\begin{eqnarray}\label{cemnonl1}
	&&(1+\alpha~ |\psi_{n}|^2) ~( 	J_L	~\psi_{n+1} +J_R ~\psi_{n-1})+\nonumber\\
	&&( \frac{g_1~|\psi_{n}|^2}{1+\beta~ |\psi_{n}|^2}+g_2~|\psi_{n}|^4+g_3~|\psi_{n}|^6 )~\psi_{n}   =E~\psi_n  
\end{eqnarray}	
where $\alpha$ and $\beta$ are either $0$ or $1$, $g_{1,2,3}$ are all positive real-valued constants, and $E$ is referred as energy for convenience. We set ${J_L=1}$ and ${J_R=\gamma}$ with $0\leq\gamma<1$ for simplicity. This equation contains five nonlinear parameters that can be chosen independently. It specifically describes a system with Kerr-type nonlinearity when ${\alpha = \beta = g_2 = g_3 = 0}$, as well as other interesting nonlinearities such as saturable nonlinearity when ${\alpha = g_2 = g_3 = 0}$ \cite{hakem0}, Ablowitz-Ladik nonlinearity when ${\beta = g_1 = g_2 = g_3 = 0}$ \cite{hakem1}, and cubic-quintic-sextic nonlinearity when ${\alpha = \beta =  0}$ \cite{hakem2}, among others. We analyze this generalized equation under the following boundary conditions: ${	\psi_{0}=\psi_{N+1}=0  }$ for open boundary conditions (OBC) and ${\psi_{N+1}=\psi_{1}}$, ${\psi_{0}=\psi_{N}}$ for periodic boundary conditions (PBC), and ${ \psi_{0}= \psi_{\infty}=0  }$ for the semi-infinite boundary conditions (SIBC). We refer to the SIBC and OBC spectra as the set of values that the parameter $E$ can take for stable skin modes localized at the left edge under SIBC and OBC, respectively.\\
Nonlinear systems exhibit two important features: chaos and fixed points. They can be explored in our system using the shooting method, which converts the boundary value problem into an initial value problem by guessing the field amplitude at the left edge. In this way, we obtain a unique solution that satisfies the boundary condition. Starting with an arbitrary initial value $\psi_1$ and $E$, we perform numerical iterations in Eq.(\ref{cemnonl1}) to compute the subsequent terms ${\{\psi_2,\psi_3,...,\psi_{N+1}\} }$, and then check if the remaining boundary condition for ${\psi_{N+1}}$ is satisfied. If it is not, the initial guess $\psi_1$ and $E$ are modified, and the numerical iteration is repeated until the solution meets the required condition for ${\psi_{N+1}}$. For bounded solutions (non-diverging $\psi_n$), our numerical analysis reveal either chaotic behavior or fixed points at high iteration values, depending on $E$. Chaos arises when a slight variation in $E$ in the numerical iterations leads to unpredictable behavior as $n$ increases. In contrast, $\psi_n$ may converge to a final value, known as a fixed point (also known as an equilibrium point). A fixed point is a point in the system where $\psi_n$ does not change with $n$ when ${n>N_c}$, where $N_c$ is the critical size. Furthermore, this fixed point is stable if small perturbations around the fixed point decay with $n$, causing the system to converge back to the fixed point.\\
To find the fixed points in our system, we require that $\psi_n$ remain unchanged when $n>N_c$, i.e., $\ds{\psi_{N_{c}+1}=\psi_{N_{c}+2}= ... =\psi_{N}=a_0 e^{i\theta}}$ in Eq.(\ref{cemnonl1}), where $a_0$ and $\theta$ are real numbers. It is worth noting that the critical size $N_c$ is a finite number at certain values of $\psi_1$ and $E$. However, determining these values as a function of the nonlinear parameters $\alpha$, $\beta$ and ${g_{1,2,3}}$ is challenging. The equation for $a_0$ reads
\begin{equation}\label{cemnonl2}
	( \frac{g_1a_0^2}{1+{\beta}~ a_0^2}+g_2 a_0^4+g_3 a_0^6 )~a_0   =(E- (1+{\alpha}~ a_0^2)~ E_{c})~a_0 
\end{equation}		
where $E_{c}=1+\gamma$. It yields two types of solutions: the zero fixed point with ${a_0 = 0}$ and nonzero ones with ${a_0\neq 0}$, which are real-valued solutions of this equation. Unlike the zero fixed point, the nonzero one is $E$-dependent and appears when ${E>E_c}$ (since ${a_0^2> 0}$). For example, $\ds{a_0^2=\frac{E-E_c}{ E_c} }$ in the case of the Ablowitz-Ladik nonlinearity, i.e., ${g_{1,2,3}=0}$ and ${\alpha=1}$. Eq.(\ref{cemnonl2}) also reveals that $\theta$ is an arbitrary parameter. In the numerical shooting method, changing the initial value ${\psi_1 \rightarrow \psi_1 e^{i\theta}}$ leaves $E$ unaffected and transforms the remaining $\psi_n$ to ${\psi_n e^{i\theta}}$. Therefore, for the sake of simplicity, we consider only real values of $\psi_1$ throughout this work.\\ 
Let us next study the stability of the fixed points. Intriguingly, the stability of the zero fixed point for the semi-infinite system is controlled by the non-Hermiticity rather than the nonlinear interactions. To be precise, the zero fixed point is stable regardless of the nonlinear interaction strengths when $E$ is enclosed by the linear PBC loop in the complex energy plane (defined by $E_{PBC} = e^{ik}+\gamma~e^{-ik}$ with $-\pi{\leq}k<\pi$). This can be demonstrated through a linear stability analysis, where a small perturbation is introduced to $\psi_n$ in Eq.(\ref{cemnonl1}) for $n{>}N_c$: ${\psi_n \rightarrow  a_0 e^{i\theta}+ \epsilon~ \phi_{n}}$ with $|\epsilon| << 1$ and ${\phi_{\infty}=0}$. By omitting terms of order $|\epsilon|^2$ and higher, we see that the resulting equation does not include any nonlinear parameters when $a_0=0$: ${\phi_{j+1}+\gamma~\phi_{j-1}=E~\phi_{j}}$ with ${j=N_c+2,N_c+3,..}$. The condition ${\phi_{\infty}=0}$ is satisfied when $E$ is inside the linear PBC loop in the complex energy plane. Beyond this energy region, the zero fixed point becomes unstable as a small perturbation does not converge to zero. This analysis also reveals another interesting result: The zero fixed point would lose its stability if the system is Hermitian $\gamma=1$, as there exists extended eigenstates $\phi_j$ in the energy interval ${[-E_c,E_c]}$. This explains why the nonlinear skin modes are unique to the non-Hermitian system with $\gamma\neq1$. Unlike the zero fixed point, the stability region of the nonzero fixed point strongly depends on the nonlinear parameters. They are typically stable within a narrow energy range, but in some particular cases, they can remain stable over an extended range of energy values. For example, we repeat the stability analysis for only the Ablowitz-Ladik nonlinearity, where ${a_0^2=\frac{E-E_c}{ E_c} }$. Assuming real valued function $\phi_j$ with real energy ${E }$, we obtain ${\phi_{j+1}+\gamma~\phi_{j-1}=E^{\prime}~\phi_{j}}$, where ${E^{\prime}=E_c(\frac{2E_c}{E}-1) }$. When ${E>E_c}$, we see that ${-E_c<E^{\prime}<E_c  }$, implying converging solutions, $\phi_{\infty}=0$. This proves the stability of the nonzero fixed point in the presence of only the Ablowitz-Ladik nonlinearity. In this paper, we consider these fixed points to construct skin and extended modes using the simple shooting method. Specifically, $a_0=0$ and $a_0\neq0$ ensure SIBC (or OBC when $N_c$ is finite) and PBC, respectively \cite{ekbilgi}. \\
We can study bifurcation diagrams for real positive values of $E$, where $E$ is treated as the bifurcation parameter. From the earlier discussion, one can get a general understanding of the bifurcation diagrams as we increase $E$ from zero. In such diagrams, $\psi_n$ converges to zero value regardless of the nonlinear parameters when ${E<E_c}$. Above this point, the zero fixed point becomes unstable, and $\psi_n$ converges to an $E$-dependent nonzero fixed value. Then, the nonzero fixed point becomes unstable at a certain value of $E$ that strongly depends on the nonlinear interaction strengths. This leads to bifurcations and ultimately to chaos (bifurcation does not appear if the nonzero fixed point is stable). As a result, in these diagrams, changing the nonlinear parameters at fixed $\gamma$ changes the points where bifurcation and chaos occur, while leaving $E_c$ unchanged. With these in mind, one can explore bifurcation diagrams with various values of the nonlinear parameters. Due to its experimental relevance, we present a bifurcation diagram for the system with only Kerr-type nonlinearity. Fig. 1(a) shows it for positive values of $E $ at ${\gamma=0.2}$ and ${g_1=1}$. The field amplitude always collapses to zero for ${E{<}E_c=1.20}$, and converges to an $E$-dependent nonzero value until $E$ reaches $2.37$, where the nonzero fixed point (period one orbit) becomes unstable, and period-2 points occur, where $\psi_n$ starts to alternate between two values instead of converging to a single number. Similarly, at a higher value of $E$, the period-2 points loose stability and bifurcate into period-4 points. This period-doubling behavior continues to period-8, period-16, and higher points, until the system exhibits chaotic behavior at ${E=2.65}$, where a tiny variation in $E$ can lead to a non-periodic behavior in the field amplitude $\psi_n$. In this case, the solutions become incompatible with the boundary conditions. The iteration eventually diverges when ${E{\geq}2.78}$.\\
Let us now construct extended and skin modes based on the bifurcation diagram in Fig. 1(a). One can obtain extended PBC modes at period-1, period-2, and higher periods until the system enters the chaotic regime. For example, at a period-2 point occurring at $E=2.37$, $\psi_j$ oscillates between two values, causing the corresponding density to exhibit a sawtooth shape, in contrast to the uniform density of the extended modes with a period of $1$. Similarly, at a period-4 point, extended modes oscillate between four distinct values, as shown in the inset of Fig. 1(a). Note that we set the initial value $\psi_1$ to either of these four values in the numerical shooting method. Consider next ${E<E_c}$ for the system under SIBC or OBC. In this case, skin modes localized at the left edge appear as $\psi_n$ converges to the zero fixed point, satisfying SIBC (and also OBC when $N_c$ is finite). As an illustration, we perform iteration for two different initial values of $\psi_1$ and plot the densities of the skin modes up to ${n=100}$ at ${E=1.0}$ (black) and ${E=1.1}$ (red) in Fig. 1(b). Note that these modes are tightly localized at the left edge even if $N_c$ is infinite. Therefore, such SIBC skin modes may practically be regarded as quasi-stationary OBC modes that maintain their forms over an extended period of time in sufficiently long lattices, as they approximately satisfy OBC \cite{nonlin9,nonlin10}. As can be seen from the figure, the localization lengths of these skin modes increase with energy $E$, similar to those observed in the linear model. However, the localization lengths depend on the power ($P=\sum_{n=1}^N|\psi_n|^2$), as illustrated in the figure, where varying $\psi_1$ from ${0.1}$ (solid) to ${0.3}$ (dashed) increases the power by a factor of nearly $7$ and $6$ for ${E=1.0}$ and ${E=1.1}$, respectively. Note that such a scaling of $\psi_1$ at fixed $E$ does not generate a distinct skin mode in the linear system. The inset of Fig. 1(b) shows the densities of the skin modes at ${\psi_1=0.5}$ and ${E=1.0}$ at four different values of $\gamma$: $0.2$ (in black), $0.7$ (in red), ${0.9}$ (in blue), and ${1.0}$ (in green). Increasing $\gamma$ extends the skin modes further from the left edge, as reduced asymmetry in the couplings weakens skin localization, leading to more extended skin modes. Notably, this extension reaches the lattice size at ${\gamma=1}$, converting skin modes into extended ones. This suggests that nonlinear skin modes are unique to the non-Hermitian system, ${\gamma\neq 1}$, which is also in agreement with our previous finding that the zero fixed point loses its stability at ${\gamma=1}$, implying the absence of skin modes. \\
\begin{figure}[t]
\includegraphics[width=4.0cm]{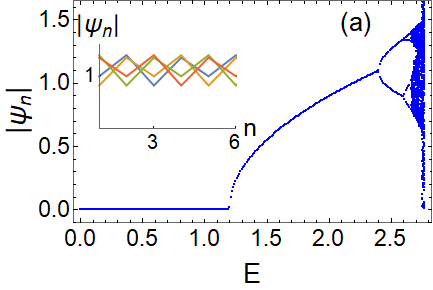}
\includegraphics[width=4.15cm]{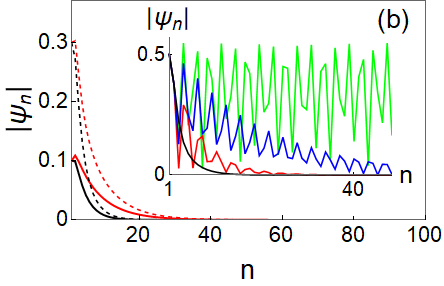}
\includegraphics[width=4.05cm]{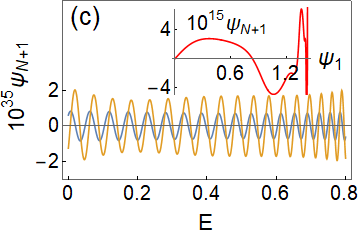}
\includegraphics[width=4.05cm]{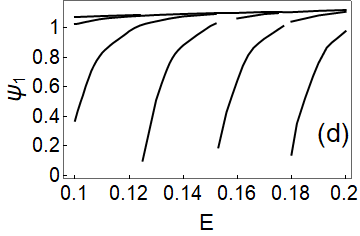}
\caption{(a) The bifurcation diagram reveals the critical value ${E_c=1.2}$, below which the zero fixed point appears, and above which $E$-dependent fixed points arise. Period doubling occurs at ${E= 2.37}$, followed by chaos at ${E = 2.65}$. In the inset, the densities of the extended PBC modes at a period-4 point with ${E=2.6}$ are illustrated up to $n=6$. (b) $|\psi_n|$ for the skin modes when ${\psi_1=0.1}$ (solid) and ${\psi_1=0.3}$ (dashed) at two different energies at $E=1.0$ (black) and ${E=1.1}$ (red). The inset demonstrates $|\psi_n|$ at ${\psi_1=0.5}$ and ${E=1.0}$ as $\gamma$ is increased from ${0.2}$ (in black) and ${0.7}$ (in red) to ${\gamma=0.9}$ (in blue) and ${1.0}$ (in green). (c) The scaled field amplitude $\psi_{N+1}$ as a function of $E$ at ${\psi_1=0.6}$ (in blue) and ${\psi_1=0.9}$ (in orange), and $N=100$. The inset shows the scaled $\psi_{N+1}$ at ${E=1.0}$ and ${N=100}$ as a function of $\psi_1$, indicating that OBC is satisfied at certain values of $\psi_1$. (d) The relation between $E$ and $\psi_1$ for OBC modes with $N=100$ in the $E$-interval [0.1,0.2]. The common parameters are ${\gamma = 0.2}$ (except for the inset in (b)), ${g_1 = 1}$ and all other nonlinear interaction strengths are zero.}
\end{figure}
A portion of SIBC skin modes are OBC skin modes. These modes exhibit some features, such as continuous spectrum, degeneracies, and energy-power branches. To explore these characteristics, one may begin by iteratively computing $\psi_{N+1}$ as a function of $E$ for an arbitrary value of $\psi_1$, and then determine the values of $E$ at which OBC is satisfied, i.e., $\psi_{N+1}(E)=0$. As an illustration, we consider Kerr-type nonlinearity with ${g_1 = 1}$ for a finite lattice with $N=100$ and ${\gamma = 0.2}$. Fig. 1(c) plots the scaled $\psi_{N+1}$ as a function of $E$ at ${\psi_1=0.6}$ (in blue) and ${\psi_1=0.9}$ (in orange). As can be seen, they oscillate as $E$ varies and touch zero at certain values of $E$. These specific values, where $\psi_{N+1}=0$, correspond to OBC skin modes at $ N=100$. This plot also shows that selecting a different arbitrary value for $\psi_1$ generates distinct values of $E$ at which $\psi_{N+1}=0$, so varying $\psi_1$ continuously generates a continuum of $E$. Note that these solutions can also be used to construct the OBC skin modes for larger lattice sizes, as $\psi_{N+1} = 0$ guarantees that $\psi_{n > N+1} = 0$ (a zero fixed point). Our numerical calculations reveal that $N_{c} $ does not take finite values when $E$ is complex, suggesting that the nonlinear OBC spectrum is real valued, similar to the linear OBC spectrum. In addition to the continuous spectrum, OBC skin modes exhibit another feature: two or more distinct OBC skin modes with different powers can share the same value of $E$. To illustrate this, we plot the scaled $\psi_{N+1}$ as a function of $\psi_1$ at ${E=1}$ in the inset of Fig. 1(c). As can be seen, the condition ${\psi_{N+1}=0}$ is satisfied at multiple values of $\psi_1$, implying degeneracy, as distinct OBC skin modes exist at the same value of $E$, each with a different $\psi_1$ (correspondingly a different power). Finally, the OBC modes display an additional feature: the energy-power branches. We perform numerical calculations to determine the relation between $E$ and $\psi_1$ when ${0.1{\leq}E{\leq}0.2}$. Fig. 1(d) reveals non-smooth behavior between $E$ and $\psi_1$. This novel type of branched structure is characterized by dramatic changes in the lowest possible values of $\psi_1$ at certain energies, implying that the powers for the OBC skin modes exhibit discontinuities in $E$. For example, an OBC mode is available at $\psi_1=0.10$ when $E=0.125$, whereas it appears at $\psi_1=0.98$ when $E$ is just below this value.  \\
The SIBC spectra for both the linear and nonlinear systems fill the same disk, bounded by the linear PBC loop in the complex energy plane, but the corresponding OBC spectra do not share this similarity. For the linear system, the OBC spectrum is confined to the real energy interval $\mp2\sqrt{\gamma}$, indicating that it is a subset of the corresponding SIBC spectrum. However, for the nonlinear case, OBC spectrum is not bounded by the SIBC spectrum and its width is size-dependent. In other words, nonlinear OBC skin modes exist at real energy values $E$ outside the range of ${\mp}E_c$. The key factor behind this intriguing behavior is that, for ${E> E_c}$, $\psi_n$ may initially decay and become zero at a certain point (satisfying OBC) before eventually rising to the nonzero fixed point (breaking SIBC). In other words, if ${\psi_{N+1}=0}$, then ${\psi_{\infty}=0}$ for the linear system. However, in the nonlinear system, it is possible for ${\psi_{\infty}}$ to be a nonzero fixed point, even when 
${\psi_{N+1}=0}$. Fig. 2 illustrates this for the system with only Kerr-type nonlinearity, while the insets correspond to the system with different values of the nonlinear parameters. Fig. 2(a) shows $\psi_{N+1}$ as a function of $\psi_1$ at ${E=1.3 > E_c}$ for three different system sizes. We see that the condition ${\psi_{N+1}=0}$ is satisfied at specific values of $\psi_1$, which are almost the same for the three lattice sizes (indistinguishable by the naked eye in the figure). Fig. 2(b) shows the density plots for three specific modes with the same energy but different $\psi_1$. These modes are OBC modes for $N=3$ (${\psi_{4}=0}$) but not SIBC modes, as $\psi_n$ eventually reaches the same fixed point. \\
\begin{figure}[t]
\includegraphics[width=4.03cm]{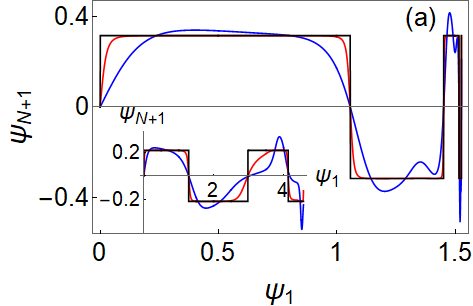}
\includegraphics[width=4.03cm]{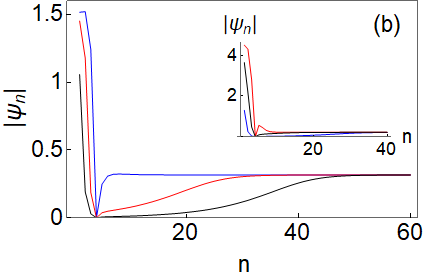}
\caption{(a) The condition $\psi_{N+1}=0$ is satisfied at specific points that are very close to each other for different lattice sizes: $N = 5$ (blue), $N = 10$ (red), and $N = 100$ (black). (b) $|\psi_n|$ for three different values of $\psi_1$ up to $n=60$, where ${\psi_n}$ becomes zero at ${n=4}$. The common parameters are ${E=1.3 > E_c}$, ${\gamma = 0.2}$, and $g_1 = 1$ (only Kerr-type nonlinearity). For the insets, the nonlinear strengths are ${\alpha = \beta = g_1 = 20g_2 = 1}$ and ${g_3 = 0}$.}
\end{figure}
\begin{figure}[t]
\includegraphics[width=3.8cm]{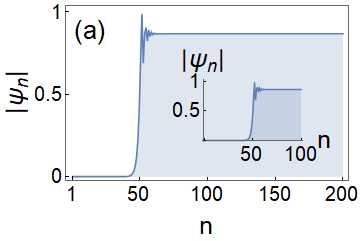}
\includegraphics[width=3.8cm]{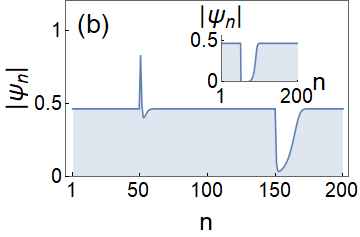}
\caption{(a) $|\psi_n|$ in the presence of a single coupling impurity at the edge when $W_1=1$, ${\psi_1=10^{-14}}$, $E=2$ and $N=200$ ($N=100$ in the inset). (b) The dark and anti-dark soliton at ${\psi_0=\psi_1= 0.461}$, $3p=p^{\prime}=150$, $W_2=-W_1=0.8$ and $E=1.4$. The width of the dark soliton can be increased by placing the second impurity next to the first one (see the inset at $p=p^{\prime}-1=50$, and $W_1=1$ and $W_2=-0.2002$). The common parameters are $\gamma=0.2$, $g_1=2g_2=5g_3=\beta=1$ and ${\alpha=0}$. }
\end{figure}
\section{ Impurity induced modes}
In the linear non-Hermitian lattice, a coupling impurity may induce an extensive number of scale-free localized states whose localization lengths scale with the system size. Here, we show that in the nonlinear case, it may lead to the formation of unique localized modes, as well as dark solitons. \\
Consider ${E > E_c}$, where $\psi_n$ converges to a nonzero fixed value. Next, we introduce a coupling impurity with strength $W_{1}$ at site ${p}$ such that $J_R$ in Eq.(\ref{cemnonl1}) is modified to ${J_R\rightarrow \gamma+W_1 \delta_{n,p}   } $ (${J_L=1}$). When the impurity with an appropriate strength is introduced at the right edge, it causes the field amplitude to drop to zero value from the nonzero fixed value at that edge. Conversely, when the impurity is located far from the right edge, it perturbs the field amplitude only locally without changing the nonzero fixed value towards the right edge. The former one may lead to localized modes under OBC, while the latter one to dark solitons under PBC, as shown below.\\
We start with the case where the impurity is located at the right edge for the lattice under OBC. Since the initial value $\psi_1$ does not change the nonzero fixed point, we propose assuming a very small value for $\psi_1$, such as ${\psi_1{<}10^{-10}}$ (${\psi_0=0}$). In this case, the field amplitude remains practically zero near the left edge until it grows sharply to a nonzero fixed value at around the point ${N_c}$, and then it remains constant towards the right edge. Note that the OBC at the right edge ${\psi_{N+1}=0}$ is satisfied with an appropriate choice of the impurity strength $W_1$. In this way, we construct novel localized modes at the right edge. As the nonzero fixed point varies with $E$, a family of such localized modes is generated by varying $E$ until $E$ becomes sufficiently large for the system to enter into a chaotic regime. The point ${N_c}$ does not change with the system size, hence these modes extend nearly from the point ${N_c}$ to the right edge. Therefore, these modes are different from the left-localized skin modes as they are localized at the right edge and their extension changes with system size. They are also different from scale-free modes localized at the right edge, which are typically given by ${\psi_n\approx e^{cx/N}}$ in the linear system, where $c$ is a constant. As an illustration, Fig. 3(a) plots $|\psi_n|$ at ${E=2}$, ${\psi_1{=}10^{-14}}$, ${W_1=1}$ when the lattice with ${N=200}$ has an saturable cubic-quintic-sextic nonlinearity. The specific lattice site where the field amplitude reaches the nonzero fixed value remains the same, regardless of the system size, as shown in the inset of Fig. 3(a) for $N=100$. As can be seen, these modes are neither skin nor scale free localized modes whose localization lengths remain constant or change proportionally with the system size, respectively.\\
A single coupling impurity may also lead to dark or anti-dark soliton solutions under PBC. A dark (anti dark) soliton is a localized drop (increase) in the density on the uniform background. To obtain them, we place the impurity far from the edges and perform the iteration with ${\psi_1=a_0}$ (period one orbit) and ${E>E_c}$ such that an extended PBC state with uniform density would be obtained in the absence of the impurity. Therefore, the field amplitude on the left side of the impurity is not changed: ${\psi_n=a_0}$ when ${n{\leq}p}$. However, it will either increase or decrease from the fixed value at ${n=p+1}$, depending on the sign of $W_1$: ${\psi_{p+1}=(1-W_1)a_0}$. Recall that the fixed point is stable against small fluctuations. Therefore the field amplitude returns to the fixed point after a few lattice sites and then remains constant towards the right edge. In this way, we obtain dark solitons when $W_1>0$ and anti-dark solitons when ${W_1<0}$. Additionally, if the system contains two well-separated impurities at sites $p$ and $p^{\prime}$ with opposite strengths, ${W_2=-W_1}$, we can generate a pair of dark and anti-dark solitons, as illustrated in Fig. 3 (b), where the impurities are located at ${p^{\prime}=3p=150}$. Introducing additional well-separated impurities into the system allows for the formation of multiple dark and anti-dark solitons. On the other hand, placing the two impurities on neighboring lattice points, $p^{\prime}=p\mp1$, allows us to adjust both the dip point and the width of the dark soliton. In the inset of Fig. 3 (b), we consider two adjacent impurities with appropriate strengths. As seen, the density drops to zero at the impurity position and then returns to its fixed point after a certain distance, leading to a wider dark soliton. Adding more neighboring impurities allows us to obtain much wider dark solitons. It is worth noting that the system also has skin modes when ${E{\leq}E_c}$, as the zero fixed point is stable against the coupling impurity. This indicates that skin modes and dark solitons coexist within the system, depending on the energy.

\section{Conclusions} 

Non-Hermitian skin effect is crucial as it reveals unique localization phenomena that has no Hermitian counterpart, and extending this effect to the nonlinear domain is essential for investigating its potential applications in advanced material science with enhanced performance, such as sensors, waveguides, or filters, where controlling or exploiting nonlinearity is advantageous. In this work, we explore a generalized nonlinear equation and show that zero fixed points can provide a framework for deriving nonlinear skin modes. We also discuss that the zero fixed point loses its stability and the nonlinear skin modes disappear when the system has no asymmetrical couplings, i.e., when the system is Hermitian. It is intriguing to find that the spectra of both linear and nonlinear skin modes under SIBC occupy the same disk, bounded by the linear PBC loop in the complex energy plane. However, the spectral region of the OBC skin modes is significantly altered by the nonlinear interaction. Furthermore, the OBC spectrum is not a subset of the corresponding SIBC spectrum. The nonlinear OBC skin modes are power-dependent, and a continuum of energies associated with OBC skin modes appear, in contrast to the discrete energies for the linear system. Additionally, two or more distinct nonlinear OBC skin modes can share the same energy. Introducing a single coupling impurity may lead to dark or anti dark soliton solutions under PBC, and localized modes that are neither skin nor scale free localized modes under OBC.  \\
This study was supported by Scientific and Technological Research Council of Turkey (TUBITAK) under the grant number 124F100. The authors thank to TUBITAK for their supports.


\begin{thebibliography}{0}
\bibitem{ekbilgi} $|\psi_n|$, rather than $\psi_n$ itself, can converge to a fixed value. However, this approach doesn't provide insights into the behavior of nonlinear skin modes, but rather addresses extended modes.
\end{thebibliography}
\end{document}